\documentclass[fleqn,10pt]{wlscirep}

\usepackage{breakurl}
\usepackage{adjustbox}
\usepackage{graphicx}

\newcommand{\given}{\operatorname{|}}

\title{Bayesian Networks Analysis of Malocclusion Data}

\author[1]{Marco Scutari}
\author[2]{Pietro Auconi} 
\author[3,4,5,*]{Guido Caldarelli}
\author[6,7]{Lorenzo Franchi} 
\affil[1]{Department of Statistics, University of Oxford, 24-29 St Giles' Oxford OX1 3LB, UK}
\affil[2]{Private Practice of Orthodontics, Roma, Italy}
\affil[3]{IMT School for Advanced Studies, Piazza San Francesco 19, 55100 Lucca, Italy}
\affil[4]{Istituto dei Sistemi Complessi CNR, Unit\`a Sapienza, Dip. Fisica , P.le A. Moro 2, 00185 Rome, Italy} 
\affil[5]{London Institute for Mathematical Sciences,  35a South St, Mayfair London W1K 2XF, UK} 
\affil[6]{Dipartimento di Ortodonzia, Universit\`a di Firenze, Firenze, Italy}
\affil[7]{Center for Human Growth and Development, University of Michigan, Ann Arbor MI, USA}
\affil[*]{Guido.Caldarelli@imtlucca.it}
\keywords{Orthodontics, Complex Networks, Community detection}

\begin{abstract}
In this paper we use Bayesian networks to determine and visualise the interactions among various
Class III malocclusion maxillofacial features during growth and treatment. We start from a sample
of 143 patients characterised through a series of a maximum of 21 different craniofacial features.
We estimate a network model from these data and we test its consistency by verifying some commonly
accepted hypotheses on the evolution of these disharmonies by means of Bayesian statistics. We
show that untreated subjects develop different Class III craniofacial growth patterns as compared
to patients submitted to orthodontic treatment with rapid maxillary expansion and facemask
therapy. Among treated patients the \textit{CoA} segment (the maxillary length) and the
\textit{ANB} angle (the antero-posterior relation of the maxilla to the mandible) seem to be the
skeletal subspaces that receive the main effect of the treatment.
\end{abstract}

\begin{document}
\flushbottom
\maketitle
\thispagestyle{empty}

\section*{Introduction} 

The use of statistical methods in medicine is crucial to overcome the large individual variability
in the pathological features of different patients\cite{scotch2010use}. In the orthodontic
discipline, the variability of craniofacial disharmonies is especially relevant due to important
differences between individuals in the amount and direction of facial growth due to heredity,
gender, ethnic background, and functional characteristics\cite{nanda2016atlas}. In this paper we
introduce an approach based on modern techniques from Bayesian statistics for Complex Network
analysis to estimate and describe the evolution of orthodontic features measured simultaneously
on a set of patients.
An incredibly large amount of integrations of the various components of the craniomaxillary and
mandibular combinations are possible during the growth process: the integration of these features
determines the ultimate dentofacial harmony or disharmony\cite{zaidel2005face}. An in-depth
understanding of the resulting large amounts of interrelated data obtained from clinical,
radiographic, and functional analyses is required to establish a solid knowledge basis for
orthodontic diagnoses. Malocclusions are isoforms of disharmony: they express a form of organic
integrity during the growth process by assimilating existing elements in a new synthesis. These
isoforms incur costs in terms of weakness of mechanotrasduction, cumulative occlusal trauma,
adaptability, local optimisation, competition between tooth elements for space, and outcome
uncertainty about the ultimate facial appearance\cite{baccetti2007growth}. 
These conditions are rarely a consequence of an abnormality in a single craniofacial component,
so individual clinical and radiological measurements are likely to be less indicative than the 
interplay between the measurements themselves. In the case of patients affected by Class III
malocclusion (characterised by the protrusion of lower dental arch), skeletal imbalance is
established early in life, becomes more pronounced during puberty, and continues to increase
until skeletal maturation is complete\cite{baccetti2007growth}. Therefore, predicting treatment
success or failure early in a single Class III patient based on a small number of morphometric
determinants is problematic\cite{fudalej2011prediction}.

Here we present a methodology that makes use of longitudinal data collected from a sample of
orthodontic patients to evaluate possible causal paths linking orthodontical features during the
growth process and the changes in those paths induced by the treatment. Practising orthodontists
often perform clinical reasoning under uncertainty about facial growth, with incomplete
information, receiving far more inputs than they can consciously consider; and as a result they
are forced to distil clinical and/or radiological evidence into regularities and
patterns\cite{silver2015signal}. Modern techniques in computational statistics build on
fundamental principles of probability theory\cite{cumming2014new} to provide a better understanding
and visualisation of complex data by learning those regularities and patterns directly from the
data, thus producing rigorous yet tractable models of domains in which expensive computations are
required for quantitative reasoning\cite{pearl1988probabilistic}. In particular, Bayesian statistics
\cite{bayes1763essay,laplace1774memoir} develops the idea of combining the information contained in
experimental data with prior knowledge available from the literature and from previous experiments
to evaluate the probability of specific hypotheses; an approach that is natural and especially
useful in biological and medical research \cite{berry,useR10}. These computational tools can
summarise a biological system involving multiple interacting components into a simplified
representation that captures the interplay between those components; and that can provide insights
on how those components influence (and possibly are causal for) each other\cite{causality}. A
convenient device to represent such complex patterns of relationships are complex networks
\cite{caldarelli2007scale}, which provide a high-level, abstracted view of the interplay between
the variables of interest by representing them as nodes and by linking them with arcs that show
how those variables interact with each other. A popular choice for this kind of representation
are Directed Acyclic Graphs (DAGs), in which links represent direct probabilistic dependencies
and have arrows indicating the direction of the dependence. 
In this paper we intend to develop such a model in the context of orthodontics, combining DAGs
with the joint probability distribution of the craniofacial variables of interest\cite{jss09}.
The contribution of the DAG in this case is to visualise the set of relationships between these
variables and to determine how they may be grouped into communities.
Networks have already been used in the literature to describe the evolution of patients with
malocclusions\cite{scala14} and to help in the formulation of diagnosis\cite{auconi11}. Indeed,
the exact focal morphological areas of the treatment effect and treatment priorities are both
still under discussion in the clinical orthodontics community. 

Unfortunately, current clinical evidence has been unable to fully elucidate the
network of causalities that link the relevant skeletal components, the starting point of the
treatment, treatment priorities, and the best way to channel and disseminate the effects of
the treatment \cite{auconi11}.
Here we will try to address such questions for Class III malocclusion, a dysmorphosis characterised
by growth excess of the mandible and/or a defective growth of the maxilla, with protrusion of the
lower dental arch. To this end we will use a set of $147$ longitudinal measurements of various
craniofacial features on $143$ Class III growing patients evaluated at least twice between the ages
of $6$ and $19$. Sixty-six of these subjects were undergoing orthodontic treatment by early rapid
maxillary expansion and facemask therapy followed by fixed appliances, while the remaining $77$
were not subject to any treatment.
We will estimate Bayesian networks from these data and we will use resampling techniques from modern
statistics to produce a consensus network model that describes the relationships between treatment
and craniofacial figures and to evaluate its predictive accuracy. 
We find the resulting network to be consistent with a number of key characteristics of Class III
malocclusion as known from current clinical evidence and literature, which we use to validate the
relationships we learn from the longitudinal data. Furthermore, the network displays good
predictive accuracy for the dynamics of Class III malocclusion in new patients. Finally, we use
the network to identify the focal morphological areas of the treatment effect on the basis of the
causal relationships captured by the network structure. 

\section*{Methods}

\subsection*{The Data}

The data contain longitudinal measurements on a set of $147$ Class III growing patients ($83$
female, $60$ male) evaluated at least twice between the ages of $6$ and $19$. Two sets of
simultaneous measurements at ages $T_1$ ($6$ to $19$ years, average $8 \pm 1$ years) and $T_2$ 
($5$ to $19$ years, average $15 \pm 1$ years) are available for all patients, in addition to a 
\textit{Treatment} variable identifying treated from untreated patients. For each untreated subject,
a \textit{Growth} variable indicating the prognosis as positive or negative in comparison with the
normal craniofacial progression was reported. The complete list and details for the $8$ variables 
for this data set can be found in the Supplementary Information.

\subsection*{Correlation Networks}

We represent the entire craniofacial system  as an aggregate structure of a variety of agents where
the clinical (e.g., radiographic, functional) features are the vertices of a network whose edges
are the relationships between them. To build the network we start from a measure of correlation
among the cephalometric variables $X_{a}, X_{b}$ and in particular we compute the Pearson 
correlation coefficient $r$ defined as
\begin{equation}
  r_{ab} = \frac {\sum_{i=1}^{n} \left(X_{a}(i)- \overline{X}_{a} \right)\left(X_{b}(i)-\overline{X}_{b} \right)}
            {\sqrt{\sum_{i=1}^{n}\left(X_{a}(i) - \overline{X}_{a} \right)^2} 
             \sqrt{\sum_{i=1}^{n}\left(X_{b}(i) - \overline{X}_{b} \right)^2} }
\end{equation}
where $X_{a}(i)$ is  the $i$-th value of the feature $X_{a}$ as observed in the data and
$\overline{X}_{a}$ is the arithmetic mean of the $n$ values observed for $X_{a}$, \textit{e.g.}
$\overline{X}_{a} = \left(\sum_{i = 1}^n X_{a}(i)\right)/n$. Pairs of variables with values of
$|r_{ab}|$ above the threshold of $0.4$ are linked by edges in the correlation network.

\subsection*{Bayesian Statistics}

The field of statistics provides several approaches to estimate the probability of particular events
of interest and to model the laws that govern the phenomena under investigation. For instance, we
can estimate the former with its observed, empirical frequency (\textit{frequentist statistics});
and the latter by making assumptions on the distribution of the data and estimating the values of
the parameters of the model as those that are best supported by the data (that is, having the
\textit{maximum likelihood})\cite{essentials}. A second approach is given by \textit{Bayesian
statistics}\cite{lee}, in which we also assume {\em a priori} a distribution for the parameters of
the model. That distribution is then updated based on the observed data to reflect the current
understanding of the phenomenon; the result is the \textit{posterior distribution} of the parameters
given the data. For instance, consider the probability $p(e \given c_i)$ of the occurrence of an
event $e$ we observe under one of several possible conditions $c_i, i = 1,\ldots, k$. A classic
approach in statistics is to estimate $p(e \given c_i)$ by means of its frequency (the ratio of how
many times the event is observed over the total number of measurements), and then to diagnose the
condition as that that has the largest $p(e \given c_i)$. On the other hand, Bayesian statistics 
answers a different question, what is the probability of each condition $c_i$ given the event $e$?
Using Bayes theorem, we can write
\begin{equation}
  p(c_i \given e) = \frac{p(e, c_i)}{p(e)} = \frac{p(c_i) p(e \given c_i)}{p(e)}
    = \frac{p(c_i) p(e \given c_i)}{\sum_{j = 1}^k p(c_j) p(e \given c_j)}
\end{equation}
to express the \textit{a posteriori} probability of $c_i$ as a function of the \textit{a priori}
probability $p(c_i)$ of the condition and $p(e \given c_i)$ relative that of the complete set of 
conditions. As a result, we obtain a completely specified probabilistic model we can use to test
experimental hypotheses and that can easily incorporate additional information available from 
external sources via the $p(c_i)$ terms. Importantly, this makes it possible to iteratively update
$p(c_i \given e)$ as new data becomes available by taking the current estimates of $p(c_i \given e)$
as the \textit{a priori} $p(c_i)$ for the new data to compute new, up-to-date estimates of
$p(c_i \given e)$.

\subsection*{Differential Equations Models}

Since we are interested in modelling the evolution of malocclusion and its response to treatment
over time, we will model the data using the differences of the craniofacial features between
different time points instead of the point raw measurements. We assume that each difference can be
modelled with a linear regression\cite{weisberg} of the form
\begin{equation}
  \Delta Y = \mu + \Delta T \beta_1 + \Delta X_1 \beta_2 + \ldots + \varepsilon_{\Delta Y} 
\label{eq:diff1}
\end{equation}
where  $\varepsilon_{\Delta Y} \sim N(0, \sigma^2_{\Delta Y})$, $\Delta T = T_2 - T_1$ and
$\Delta Y = Y_{T_2} - Y_{T_1}$ and so forth for the other regressors. We can then rewrite
Eq.~\ref{eq:diff1} as
\begin{equation}
  \frac{\Delta Y}{\Delta T} = \mu^* + \frac{\Delta X_1}{\Delta T}\beta_2^* + 
    \ldots + \varepsilon_{\frac{\Delta Y}{\Delta T}}
\label{eq:diff2}
\end{equation}
which in the limit of $\Delta T \rightarrow 0$ can be considered as a set of differential equations
that models the rates of change. (This is a particular case of structural equation models\cite{sem},
which are widely used in statistical genetics and systems biology\cite{sem2,sem3}.) The relationships
between the differences are assumed to be well approximated by a linear behaviour. This constraint
is intrinsically enforced by the data: only $120$ out of $147$ patients have been measured only
twice, making it impossible to estimate any trend more complex than linear. Note that Eq.~\ref{eq:diff1}
and \ref{eq:diff2} imply that craniofacial features change linearly over time,
because each rate of change $\Delta Y / \Delta T$ depends on the rates of change of other variables
but not on time itself. To have a nonlinear trend we would need
\begin{align}
  &\Delta Y = \mu + \Delta T \beta_1 + (\Delta T)^2 \beta_2 + \ldots&
  &\Longrightarrow&
  &\frac{\Delta Y}{\Delta T} = \mu^* + \Delta T \beta_2^* + \ldots&
  &\Longrightarrow&
  &\frac{\Delta Y}{\Delta T^2} = \beta_2^* \neq 0.
\end{align}
Furthermore, including the \textit{Growth} and \textit{Treatment} in the differential equations
makes it possible to have regression models of the form
\begin{equation}
  \frac{\Delta Y}{\Delta T} = \mu^* + \frac{\mathrm{Growth}}{\Delta T}\beta_G^* +
    \frac{\mathrm{Treatment}}{\Delta T}\beta_{TR}^* + \frac{\Delta X_1}{\Delta T}\beta_2^* + 
    \ldots + \varepsilon_{\frac{\Delta Y}{\Delta T}}
\end{equation}
thus allowing for different rates of change depending on whether the patient shows positive 
developments or not in the malocclusion and whether he is being treated or not. Conversely, we do
not allow the treatment level to depend on $\Delta T$, since patients are either treated or
untreated for the whole period of observation; and we do not assign a regression model to
$\Delta T$ because we assume that it does depend on any measured variables.

\subsection*{Bayesian Networks}

A Bayesian network\cite{koller,crc13} is a statistical model to describe probabilistic
relationships among a set of variables using a directed acyclic graph (DAG). The global
distribution of the variables $\mathbf{X} = \{ X_1, \ldots, X_N \}$, where $N$ is the number of
different features (in this case $N=8$) is decomposed into a the local distributions of the
individual variables $X_i$ as 
\begin{equation}
  p(\mathbf{X}) = \prod_{i=1}^N p(X_i \given Pa(X_i))
\end{equation}
where $Pa(X_i)$ are the variables that correspond to the parents of $X_i$ in the DAG
(i.e. the nodes with an arc pointing towards $X_i$). The process of estimating such model is called
{\em learning}, and consists in two steps: 
\begin{itemize} 
  \item ``learning'' which arcs are present in the graph (i.e. which probabilistic relationships are
    supported by the data); 
  \item ``learning'' the parameters that regulate the strength of those dependencies. 
\end{itemize}
The former is known as {\em structure learning}, and the latter as {\em parameter learning}. In the
context of the differential equations described above in Eq.~\ref{eq:diff1}, in structure learning
we determine which regressors (if any) are present in each differential, while in parameter
learning we estimate the values of the corresponding regression coefficients. In order to do that,
we assume that the errors in each differential equation (represented by the $\varepsilon_{\Delta Y}$
term) are normally distributed, independent, homoscedastic and with mean zero. Under these
assumptions, each differential equation can be treated as a classic linear regression model and
estimated by ordinary least squares\cite{draper}; and the regressors correspond to the variables
associated to the nodes that are parents of $\Delta Y$ in the DAG.

Structure learning is similarly based on model selection procedures for classic regression models. 
Since we operate in a Bayesian setting, we select which variables are statistically significant
regressors in each differential equation as those that maximise the posterior probability of the
Bayesian network, which we approximate with the {\em Bayesian Information 
Criterion}\cite{schwarz}. Those regressors are the parents of the node corresponding to the
response variable in the DAG, and are chosen using hill-climbing, a greedy search algorithm
based on step-wise selection\cite{norvig}. The only restriction imposed by Bayesian networks is that,
once the probabilistic relationships are represented as a directed graph, the graph should be
acyclic.

While it is possible in principle to learn all dependencies from the data, Bayesian networks can
easily include prior knowledge available from the literature and the practice of the discipline to
produce more informative models and to overcome the inherent noisiness of orthodontic data. This
can be done by encoding the available prior knowledge in sets of \textit{whitelisted arcs} (which
we know represent real dependencies and thus should be forced to be present in the graph) and
\textit{blacklisted arcs} (which correspond to relationships we know to be impossible). In
particular:
\begin{itemize}
  \item Craniofacial features do not determine $\Delta T$ or \textit{Treatment}, so we blacklist
    any arc from the former to the latter. We also blacklist any arc from the craniofacial features
    to \textit{Growth}, as we interpret them to be determined by the overall evolution of the
    malocclusion (including unobserved factors) as expressed by \textit{Growth}. This also leads
    to a more intuitive parameterisation of the differential equations, with different regimes for
    the craniofacial features depending on the prognosis.
  \item We blacklist any arc from $\Delta T$ and \textit{Treatment} as discussed above.
  \item We whitelist the dependence structure $\Delta ANB \rightarrow \Delta IMPA \leftarrow 
    \Delta PPPM$ \cite{scotch2010use,baccetti2007growth,fudalej2011prediction}.
  \item We whitelist the arc from $\Delta T$ to \textit{Growth} to allow the prognosis to change
    over time.
\end{itemize}

We want to clarify the meaning of the variable \textit{Growth} we use in our analysis:
\textit{Growth} reflects the expected prognosis of the patient at the time of the visit. In such
respect this is a ``static'' variable and does not reflect a measure of the effective growth.
Quality of growth has been evaluated by considering the normal evolution of the  maxillomandibular
sagittal imbalance (CoGn-CoA) with respect to average population. Patients near the average
values were diagnosed as ``good growers'' while the others were indicated as ``bad growers''
\cite{auconi17}. Furthermore, to reduce the impact of the noise present in the data, we use a
second Bayesian technique called {\em model averaging} to improve the reliability of structure
learning\cite{aime11}. Typically, to examine the phenomenon under investigation we estimate a
single model from the data, and we draw our conclusions from that model treating it as a ``fixed''
quantity. In doing so we underestimate the degree of uncertainty present in those statistical
conclusions by ignoring the fact that the estimated model is not ``fixed'', but carries its own
uncertainty from the selection procedure used to learn it from the data\cite{claeskens}.
Intuitively, we can imagine that adding or removing a few observations from the data may result in
a different model being identified, in turn leading to different conclusions. To reduce this model
uncertainty, we re-sample the data $200$ times using bootstrap\cite{efron} and we perform structure
learning separately on each of the resulting samples, thus collecting $200$ DAGs. We then compute
the frequency with which each appears in those $200$ graphs, known as the \textit{arc strength},
and we compute an ``average'', consensus DAG by selecting those arcs that have a frequency above a
certain threshold. (The threshold can either be estimated from the data or set to an arbitrary
value, such as $0.85$ below, for the purpose of obtaining a sparse DAG that is easier to interpret.)
The averaged Bayesian network model has a number of favourable statistical properties; in particular,
it is less sensitive to noisy data and typically produces more accurate predictions for new
observations.

Once we have estimated the average Bayesian network and the values of the regression coefficients
in the differential equations it describes, we evaluate its predictive accuracy using 10-fold 
cross-validation\cite{elemstatlearn}. $10$-fold cross-validation is a model validation technique
that assesses how well a statistical model generalises to independent data or, in other words, how
accurately it will predict the behaviour of new observations. It is implemented as follows.
\begin{enumerate}
  \item We split the data into 10 subsets (called \textit{folds}) of the same size (or as close as
    possible).
  \item For each fold in turn:
  \begin{enumerate}
    \item we take that fold as the \textit{test set};
    \item we take the rest of the data as the \textit{training set};
    \item we learn the Bayesian network model on the training set, both the structure and the
      parameters;
    \item we predict each variable in turn for the observations in the test set, from the model we
      learned from the training set and from all the other variables in the test set;
  \end{enumerate}
  \item We collect the pairs of (observed, predicted) values for all the observations and:
  \begin{enumerate}
    \item for each continuous variable, we compute the correlation between the observed and
      predicted pairs (this quantity is called \textit{predictive correlation});
    \item for Growth, we compute the number of misclassified predicted values using the observed
      values as the true values (this is the \textit{predictive classification error}, which the
      complement of predictive accuracy).
  \end{enumerate}
\end{enumerate}

In addition, we use the averaged Bayesian network for inference to check answer a number of crucial
questions and to check whether it reflects the available knowledge of how the measured variables
interact with each other and with the treatment. In the context of Bayesian networks, this is
typically done using a technique called {\em belief updating}, in which we estimate the posterior
probability of a certain event or the posterior estimate of some parameter conditional on some
evidence on the values of one or more variables. Several exact and approximate approaches are
available in the literature\cite{crc13}; in this paper we use logic sampling for its simplicity.
Logic sampling is defined as follows:
\begin{enumerate}
  \item Define a counter $n_\mathbf{E} = 0$ for the evidence and a counter $n_\mathbf{q} = 0$ for
    the event of interest;
  \item For a suitably large number of times ($10^4$--$10^6$):
  \begin{enumerate}
    \item generate a random sample from the Bayesian network.
    \item if the random sample matches the evidence, set $n_\mathbf{E} = n_\mathbf{E} + 1$;
    \item if the random sample matched both the evidence and the event,
      set $n_{\mathbf{E}, \mathbf{q}} = n_{\mathbf{E}, \mathbf{q}} + 1$.
  \end{enumerate}
  \item Estimate the conditional probability of the event given the evidence with
    $n_{\mathbf{E}, \mathbf{q}} / n_\mathbf{E}$; or compute the posterior estimates of the
    parameters of interest using those random observations that match the event.
\end{enumerate}
By applying this approach we can answer arbitrary questions, which are called {\em conditional
probability queries}, from a Bayesian network. We perform the whole analysis using the bnlearn
package\cite{jss09} for R\cite{rcore}.
 
\section*{Results}

\subsection*{Raw data}

\begin{figure}
  \centering
  \includegraphics[height=0.35\textheight]{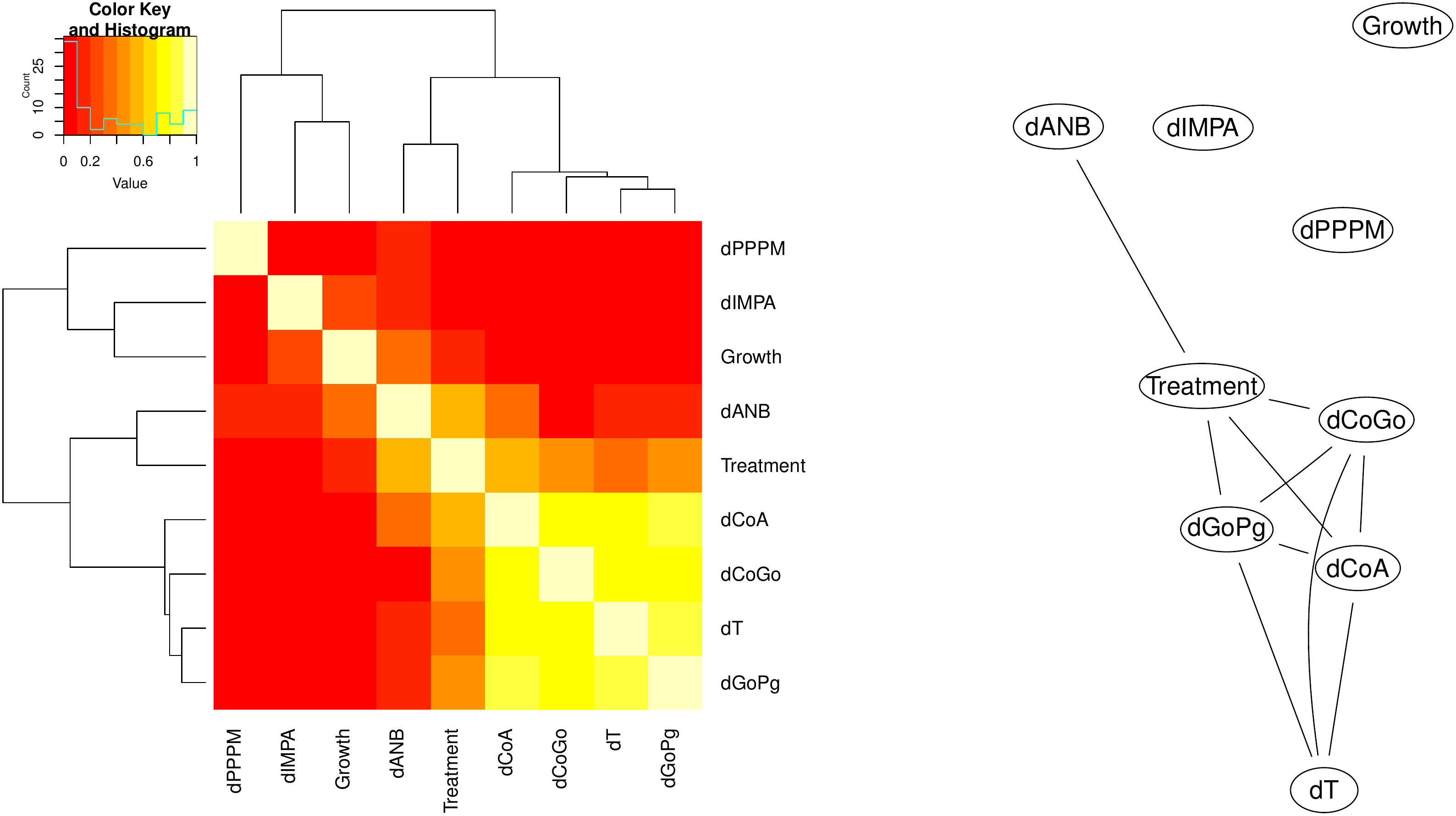}% was {correlation-heatmap3-(1,2).pdf} 
  \caption{(left) A heatmap of the Pearson's correlation between variables measured on the
    patients. (right) The correlation network displaying Pearson's correlations greater 
    than $0.40$ in absolute value.} 
  \label{heatmap1}
\end{figure}

As discussed in the literature\cite{auconi11,scala14}, craniofacial features evolve and respond to
external stimuli as a system and therefore they form clusters of variables with high (greater than
$0.40$ in absolute value) Pearson's correlation. We observe this phenomenon in Fig.~\ref{heatmap1}
(right panel), in which all the craniofacial features are connected with the exception of
$\Delta IMPA$ and $\Delta PPPM$. \textit{Treatment} is also connected to four of these features,
and presents similar correlations values with all of them (Fig.~\ref{heatmap1}, left panel), making
it difficult to establish the focal point of its action.

\begin{figure}
  \centering
  \includegraphics[width=0.55\textwidth]{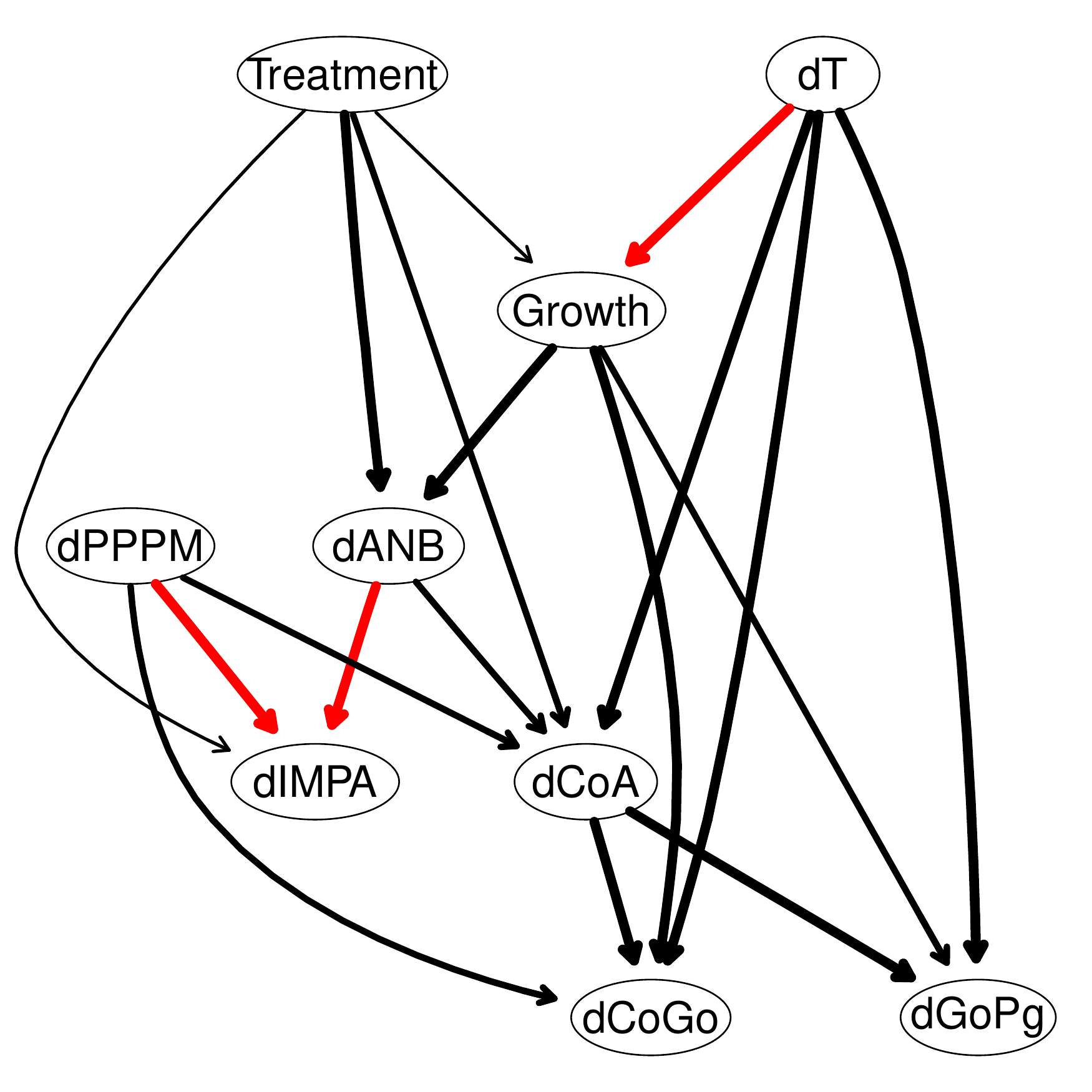}% was {figures/raw-plotting-1.pdf} 
  \caption{The DAG underlying the consensus Bayesian network learned from the variables measured on
    all $143$ patients. Arcs in red are constrained to be present in the network by the whitelist.
    The thickness of the arcs is in the proportion to their strength; only arcs with a strength
    greater than $0.5$ are included in the consensus network.} 
  \label{bn-raw}
\end{figure}

\begin{figure}
  \centering
  \includegraphics[width=0.55\textwidth]{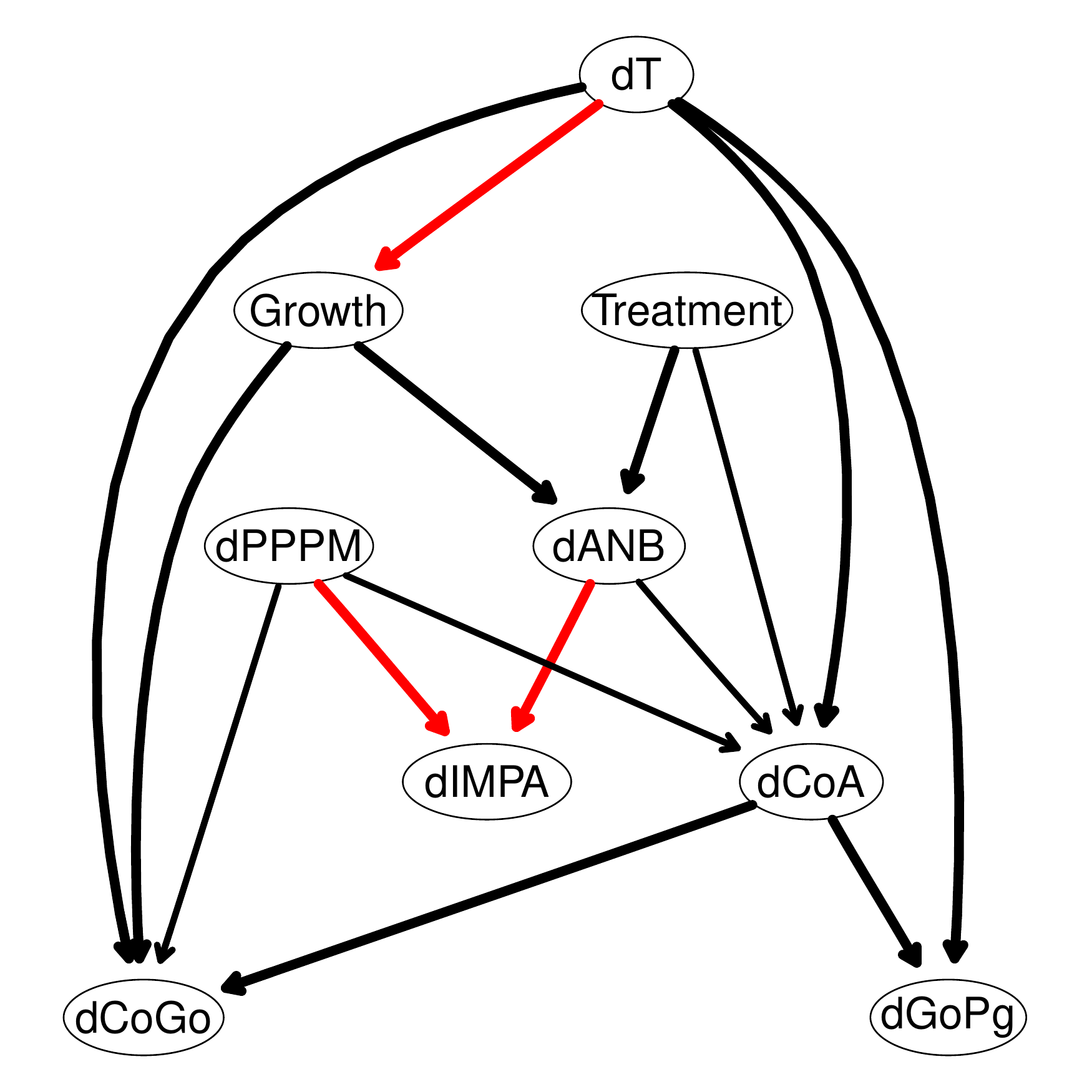}% was{figures/definitely-simpler-1.pdf} 
  \caption{A simplified DAG derived from that in Fig.~\ref{bn-raw} after removing arcs with a
     strength smaller than $0.85$.} 
  \label{simple-raw}
\end{figure}

The Bayesian network consensus model constructed by learning $200$ networks from the data and 
keeping the arcs that appear at least $50\%$ of the time (threshold estimated from the data) is
shown in Fig.~\ref{bn-raw}. All the directions of the arcs seem to be well established; this can 
probably be attributed to the use of a whitelist and a blacklist, as they force the directions of
nearby arcs to cascade into place. Furthermore, a cursory examination of the arc strengths above the
threshold confirms that $15$ out of $18$ arcs in the consensus network appear in fact with a
frequency of at least $0.85$. All arc directions are also clearly established (all frequencies are
equal to $1$). This allows to further simplify the consensus network as shown in
Fig.~\ref{simple-raw} while losing little information in the process.
While the skeletal growth process influence the evolution of sagittal maxillomandibular imbalance
($\Delta ANB$) and mandible ramps height ($\Delta CoGo$), the treatment effects mainly influence
the maxillary length growth ($\Delta CoA$), and the progression of maxillomandibular imbalance
($\Delta ANB$).

To further validate the Bayesian network, we check whether this is consistent with prior information
on Class III malocclusion that has not been used in the construction of the model. We formalise this
prior information into four hypotheses, and we use conditional probability queries as described in
the Methods (with $10^4$ random samples) to test them.

\begin{figure}
  \centering
  \includegraphics[width=0.55\textwidth]{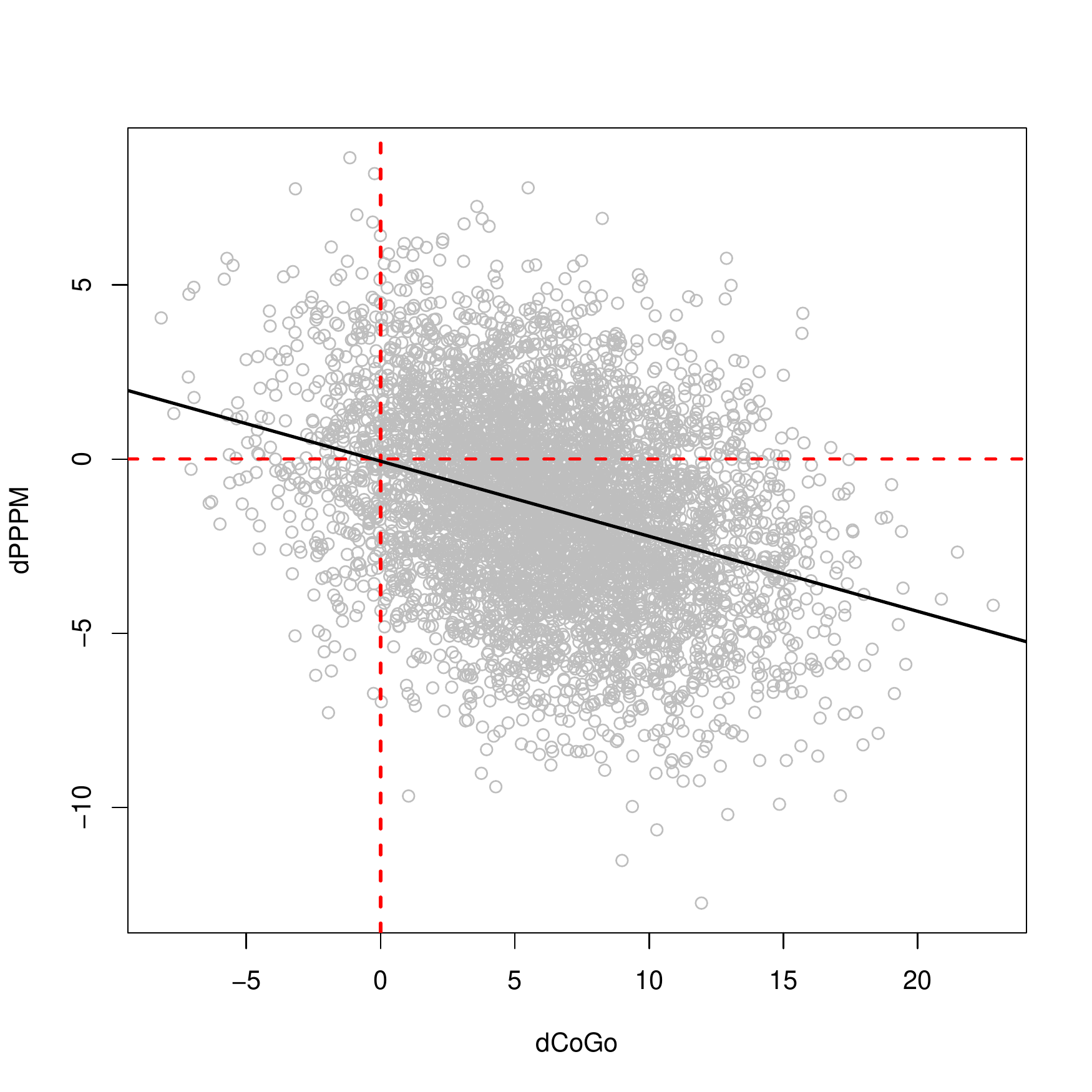}% was {figures/question1-1.pdf} 
  \caption{Values simulated from the Bayesian network for $\Delta PPPM$ and $\Delta CoGo$. The
    black line represents the regression line of $\Delta PPPM$ against $\Delta CoGo$; its
    negative slope confirms that as $\Delta CoGo$ increases (which indicates an increasingly
    rapid growth) $\Delta PPPM$ becomes increasingly negative (which indicates a
    reduction in the angle).} 
  \label{fighp1}
\end{figure}

\begin{figure}
  \centering
  \includegraphics[width=0.55\textwidth]{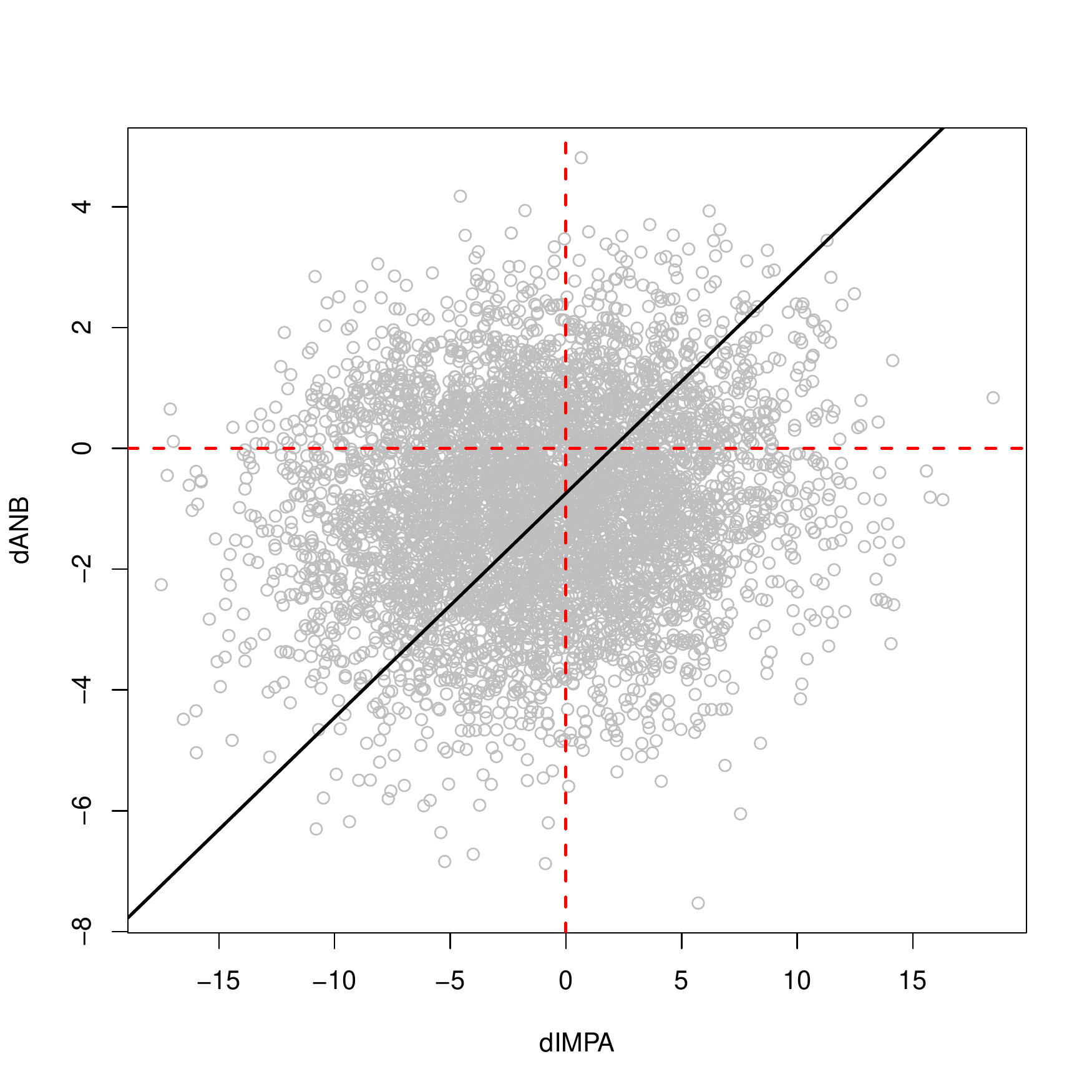}% was {figures/question3-1.pdf} 
  \caption{Values simulated from the Bayesian network for $\Delta ANB$ and $\Delta IMPA$. The
    black line represents the regression line of $\Delta ANB$ against $\Delta IMPA$; its positive
    slope suggests that $\Delta ANB$ is proportional to $\Delta IMPA$, so a decrease in one 
    suggests a decrease in the other.} 
  \label{fighp2}
\end{figure}

\begin{enumerate}
  \item In Class III growing subjects an excessive growth of \textit{CoGo} induces a reduction in
    \textit{PPPM}, assuming no treatment is taking place. In the differential equations in the
    network, we have that as $\Delta CoGo$ increases (which indicates an increasingly rapid growth)
    $\Delta PPPM$ becomes increasingly negative (which indicates a reduction in the angle). This
    is shown in Fig.~\ref{fighp1}.
  \item In Class III growing subjects, if \textit{ANB} decreases \textit{IMPA} decreases to
    compensate; $\Delta ANB$ is proportional to $\Delta IMPA$, as shown in Fig.~\ref{fighp2}, so a
    decrease in one suggests a decrease in the other.
  \item Since Class III orthodontic treatment is aimed at stopping the decrease of
    \textit{ANB} ($\Delta ANB \approx 0$), we expect to observe different dynamics for \textit{ANB}
    in treated and untreated patients. First, we note that the Bayesian network correctly 
    assigns a higher probability of a favourable prognosis to treated ($0.63$) compared to untreated
    ($0.51$) patients. The unfavourable prognosis of treated patients was defined as the concurred
    presence of Class III permanent molar relationship and negative overjet \cite{auconi2014prediction}.
    If we simulate the treatment effect and fix $\Delta ANB \approx 0$ (thus making it independent
    from its parents and removing the corresponding arcs), we have that the probability of a
    favourable prognosis is the same ($0.58$) for both treated and untreated patients and thus it
    does not depend on the treatment. This suggests that a favourable prognosis of a Class III
    malocclusion is determined mainly by preventing changes in \textit{ANB}. 
  \item If we use \textit{GoPg} as a proxy for point B, the treatment does not affect point B after
     controlling for point A: if we keep \textit{GoPg} fixed ($\Delta GoPg \approx 0$) the angle
     between point A and point B ($\Delta ANB$) evolves differently for treated and untreated
     patients. On average, $\Delta ANB$ increases for treated patients ($+0.37$ degrees; strongly
     negative values denote horizontal imbalance, so a positive rate of changes indicate a
     reduction in imbalance) and decreases for untreated patients ($-1.13$ degrees; the imbalance
     slowly worsens over time).
\end{enumerate}

Finally, we also consider the predictive accuracy of the consensus Bayesian network. Using
cross-validation as described in the Methods, we find that the prognosis is accurately predicted
with probability $0.73$. The predictive correlations for the craniofacial features are $0.86$ for
$\Delta CoGo$, $0.91$ for $\Delta GoPg$, $0.92$ for $\Delta CoA$, $0.23$ for $\Delta IMPA$, $0.42$
for $\Delta PPPM$ and $0.65$ for $\Delta ANB$.

\begin{figure}
  \centering
  \includegraphics[width=0.55\textwidth]{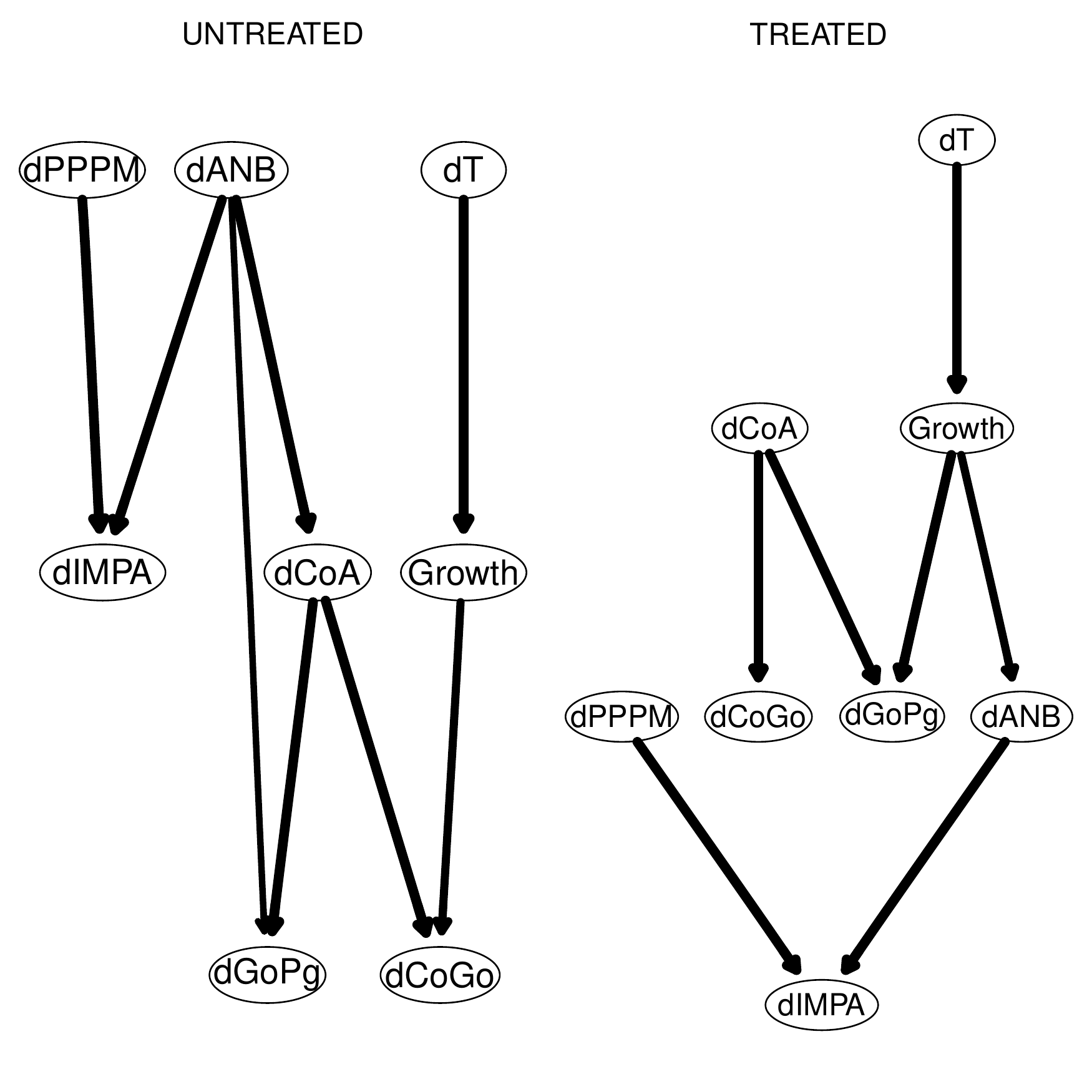}% was {figures/contrast-plot-bathia-1.pdf} 
  \caption{The DAGs underlying the consensus Bayesian networks for treated and untreated patients
     on the $8$ variables measured for both, after adjusting them using the population reference
     values from Bathia and Leighton\cite{bathia}.} 
  \label{contrast-bathia}
\end{figure}

We also learn two separate consensus Bayesian networks from treated and untreated patients, which
are shown Fig.~\ref{contrast-bathia}.
There are significant differences between the influence networks pertaining to treated and untreated
subjects. The treatment effects on the craniofacial subspaces are channelled through the maxillary
node $\Delta CoA$ to the mandibulary nodes $\Delta CoGo$ (mandibular ramus) and $\Delta GoPg$ 
(mandibular body). The adaptive $\Delta IMPA$ node (which aims to maintain unchanged the sagittal
relationship between the maxilla and the mandible) is influenced by both horizontal and vertical
skeletal imbalances during the treatment process (i.e., by $\Delta PPPM$ and $\Delta ANB$). On the
contrary, among Class III untreated subjects the progression of the horizontal skeletal imbalance 
($\Delta ANB$) strictly influences the maxillary node $\Delta CoA$, which in turn influences the
progression of mandibulary nodes $\Delta GoPg$ and $\Delta CoGo$ during the growth process.

\subsection*{Adjusted data}

\begin{figure}
  \centering
  \includegraphics[width=0.55\textwidth]{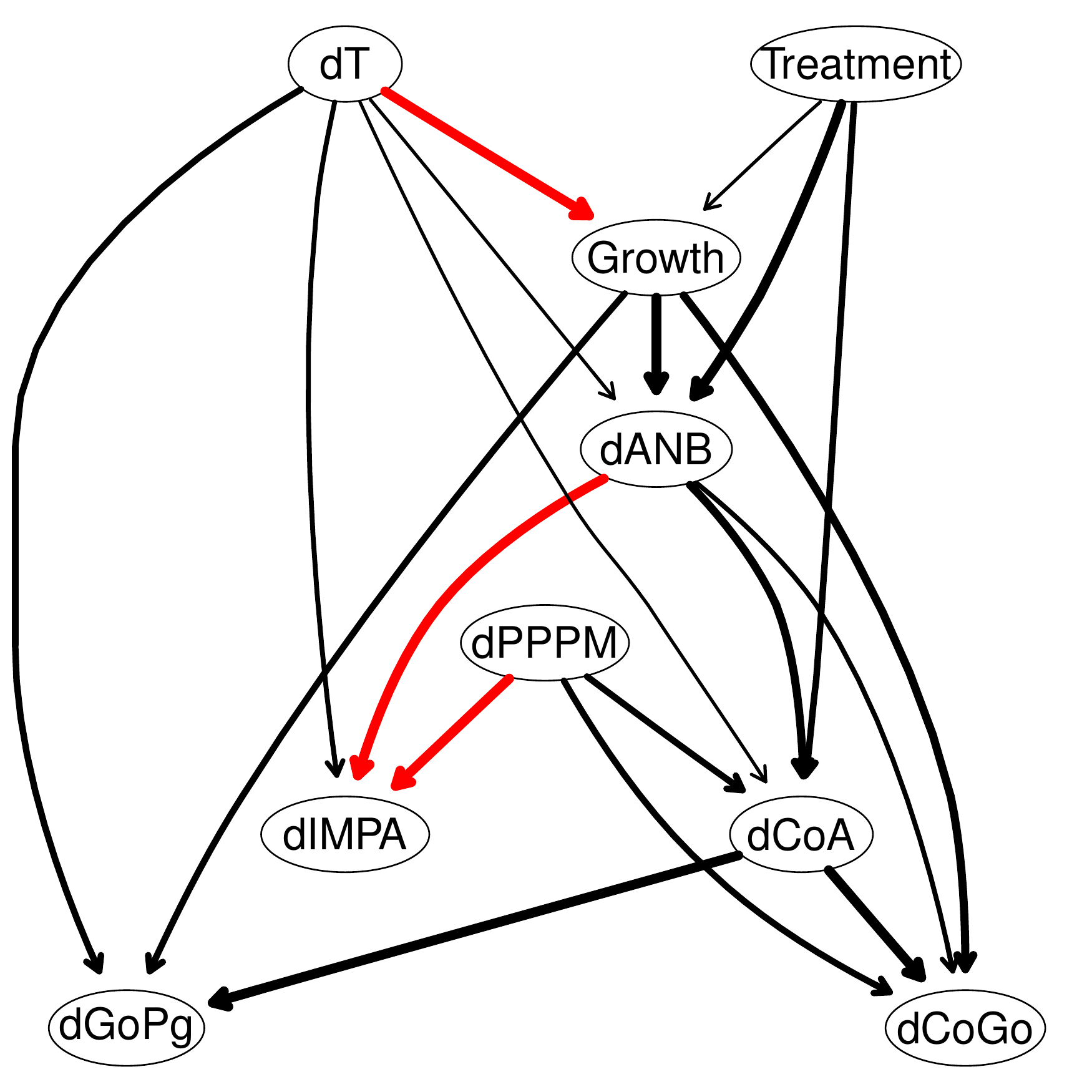}% was {figures/strength-bathia-1.pdf} 
  \caption{The DAG underlying the Bayesian network learned from the $9$ variables measured on all
     $143$ patients after adjusting them using the population reference values from Bathia and
     Leighton\cite{bathia}.} 
  \label{bn-bathia}
\end{figure}

\begin{figure}
  \centering
  \includegraphics[width=0.55\textwidth]{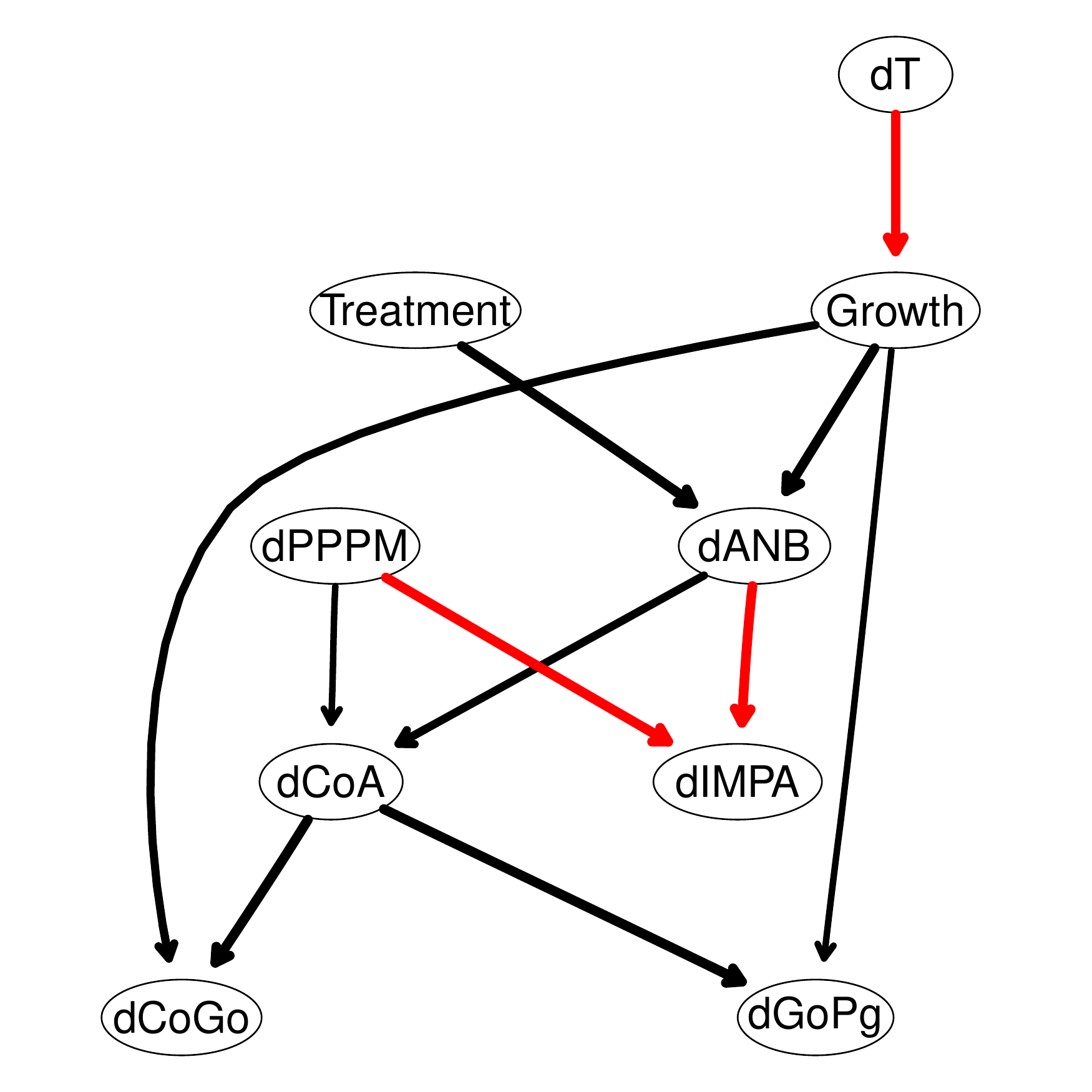}% was {figures/bathia-simpler-1.pdf}
  \caption{A simplified DAG derived from that in Fig.~\ref{bn-bathia} after removing arcs with a
     strength smaller than $0.85$.}
  \label{bn-bathia-simpler}
\end{figure}

With the aim of confirming the methodological approach we introduced in this paper, we considered
another case study. This is obtained by adjusting the data by subtracting the corresponding reference
values from an Atlas of normal cephalometric features \cite{nanda2016atlas} for infancy and 
childhood. A consensus Bayesian networks built in the same way as that for the raw data is shown in
Fig.~\ref{bn-bathia}. The threshold for the significance of the arcs is about the same as before
($0.5$) and the number of arcs is also similar. Again we can simplify the network by retaining only
the arcs with an arc strength of at least $0.85$. The most striking feature of this new network,
shown in Fig.~\ref{bn-bathia-simpler}, is the absence of arcs between $\Delta T$ and the orthodontic
variables; the only arc from $\Delta T$ points to \textit{Growth} and is only included because of
the whitelist. (Note that the same happens when we exclude the individuals with the most extreme
$\Delta T$, and the resulting networks in the two cases are very similar.) This seems to suggest that
much of the dependence on $\Delta T$ observed in the raw data is not a consequence of the evolution
of the malocclusion but a result of ageing; and it is consistent with the fact that if we reduce the
spread in the observed ages by removing the most extreme $\Delta T$, most of the dependencies on
$\Delta T$ vanish. We can conjecture that the nonlinear trend of the raw values of the orthodontic
variables (that is, the fact that that their rate of change is a function of time) can be decomposed
into two components: a general population average and a deviation from that average given the
malocclusion. The former effectively changes with time (that is, the trend of the population mean
over time is not constant) while the latter does not (that is, the rate of change of the deviation
from the population mean depends only on other orthodontic values and on the treatment). In other
words, the population average evolves with age for all orthodontic variables, which is expected as
the patients are not yet fully grown adults. However, the deviations from the populations averages
do not seem to evolve with age, or to be a function of the passage of time. This would imply that
the effects applying a change to one of the orthodontic variables propagate to related orthodontic
variables and cause them to respond them in the same way regardless of how quickly the change is
applied. (e.g. a one-degree shift in \textit{ANB} influences neighbouring variables such as 
\textit{CoA} and \textit{IMPA} in the same way regardless of how quickly that one-degree change
happens; it can be one year, it can be two years, etc. but those neighbouring variables will have
the same value at the end).

We consider the predictive accuracy of this new consensus Bayesian network, using cross-validation
as we did for the network we learned from the raw data. We find that the probability of correctly
predicting the prognosis is similar ($0.74$ vs $0.73$). The predictive correlations for most
craniofacial features, however, are smaller: $0.64$ ($-0.22$) for $\Delta CoGo$, $0.68$ ($-0.23$)
for $\Delta GoPg$, $0.81$ ($-0.11$) for $\Delta CoA$, $0.28$ ($+0.05$) for $\Delta IMPA$, $0.39$
($-0.03$) for $\Delta PPPM$ and $0.71$ ($+0.06$) for $\Delta ANB$. This is expected since we are
now modelling deviations from the population average, which are intrinsically more difficult
to analyse.

\begin{figure}
  \centering
  \includegraphics[width=0.55\textwidth]{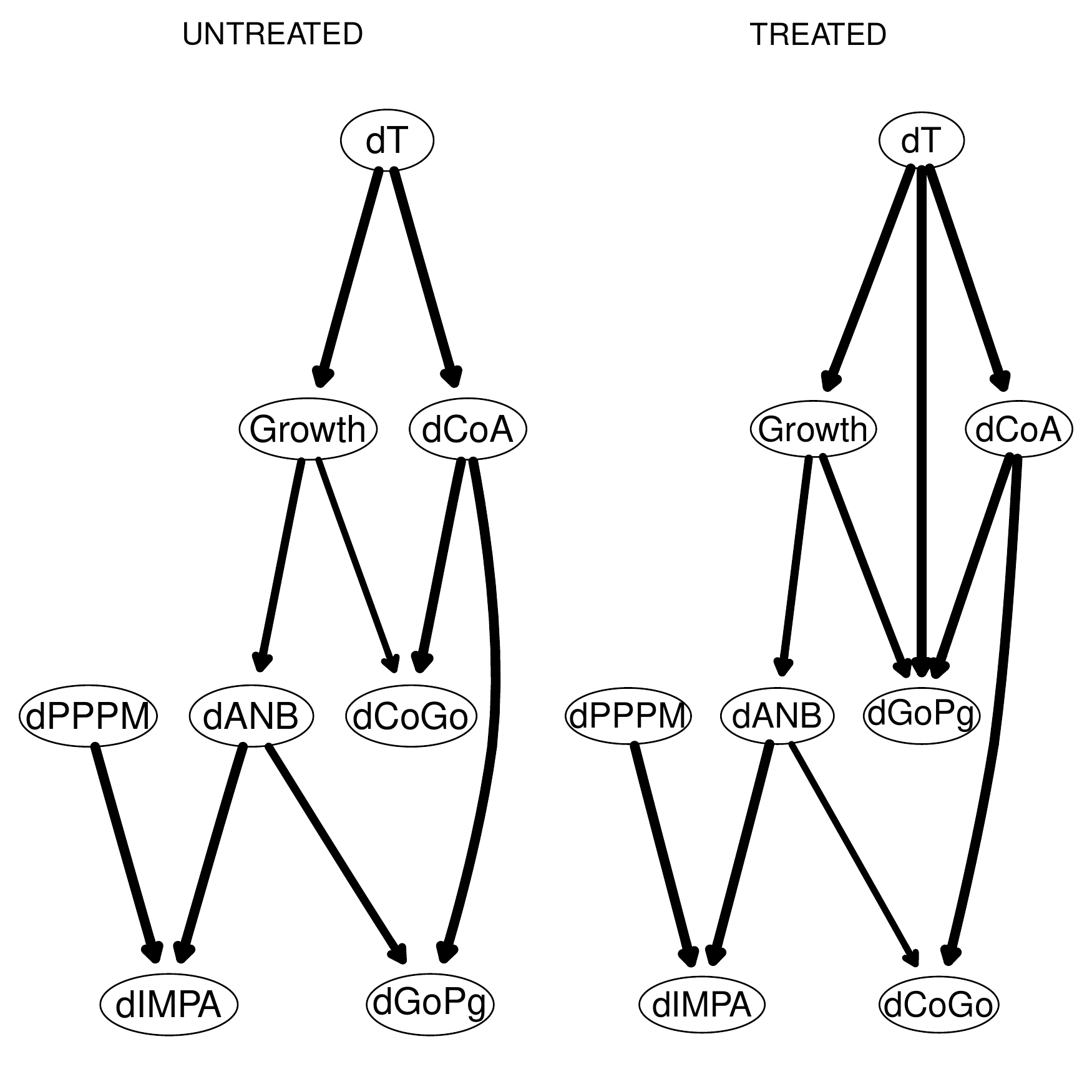}% was {figures/contrast-plot-1.pdf} 
  \caption{The DAGs underlying the consensus Bayesian networks for treated and untreated patients
     on the $8$ variables measured for both.} 
  \label{contrast-raw}
\end{figure}

We also learn two separate consensus Bayesian networks from treated and untreated patients, which
are shown Fig.~\ref{contrast-raw}. Among both untreated and treated subjects the progression of
skeletal imbalance is influenced by the evolution of the maxillary sagittal dimensions ($\Delta CoA$);
however, only treated patients exhibit the strong dependency of the mandibulary corpus from $\Delta CoA$.

\section*{Discussion} \label{sec:IV}

Previous studies have proposed different cephalometric models to determine specific facial
parameters related to abnormal growth patterns in Class III untreated and treated patients
\cite{nanda2016atlas,baccetti2007growth,fudalej2011prediction}. Multilevel, nonparametric, and
predictive function algorithms have provided growth predictions and treatment outcomes based on a
variety of facial characteristics\cite{baccetti2007growth,fudalej2011prediction}. Recently,
network approaches to understanding morphological and functional relationships among orthodontic 
data have been proposed\cite{scala14,auconi11}. While this approach improved the interpretation
of quantitative, patient-specific information, networks were unable to elucidate the effects
of influences (possibly, causal influences) between craniofacial variables\cite{scala14}. 
Craniofacial features change and adapt as a system in response to both natural stimuli such as
growth and external stimuli such as clinical treatments. This implies that efforts towards 
understanding diseases such as malocclusion must reflect this interplay in the choice of statistical
models and in how clinically relevant hypotheses are tested. This has motivated the use of network
approaches\cite{auconi11,scala14} which represent features as nodes in an undirected graph and
explicitly groups them into clusters based on their pair-wise correlation. These clusters can
describe particular regions of the craniomaxillary and mandibular complexes and other broad
features (symmetry, proportions, etc.) that are impacted by malocclusion and that must be targeted
by treatment.

This modelling approach, however, has three important limitations. The first is that the use of
pair-wise correlations makes it impossible to distinguish direct relationships between two features
from indirect ones that are mediated by other features, thus making it difficult to get a clear
picture of system as a whole and to identify the best target for the treatment\cite{opgen-rhein}.
Furthermore, the direction of the relationships is not taken into consideration by the model nor
it is represented in the graph, making it impossible to infer cause-effect relationships even in
the presence of data systematically collected from a clinical trial\cite{causality}. Finally, 
limiting the model to a representation of pair-wise dependencies falls short of characterising
the full probability distribution of the features, and that in turn makes it impossible to use
it to test the complex hypotheses required for model validation and treatment evaluation.

Bayesian networks suffer from none of these limitations\cite{useR10}. Thanks to their modular model
structure and to the availability of efficient software implementations\cite{jss09}, they can be
used to simultaneously explore a large number of features. The number of features does not impact
the interpretability of the network: focusing on direct dependencies means that each feature is
described by a local distribution that depends only those features (the ``parents'') for which the
corresponding nodes have an arc with an arrow pointing to that particular feature. Therefore, 
complex models are divided into a collection of simpler problems which are mathematically tractable
and computationally simpler. Furthermore, the DAG can always be used as a high level abstraction
for qualitative reasoning in the context of exploratory analysis and to investigate hypotheses on
the whether various sets of features are related to each other\cite{koller}. Finally, a Bayesian
network can also be interpreted as a causal network in the absence of confounding 
factors\cite{causality} and used to examine or generate novel clinical hypotheses\cite{lucas}.
The inherently Bayesian nature of a Bayesian network facilitates such reasoning by incorporating
both prior knowledge about the variables of interest and the uncertainty present in the 
data\cite{mukherjee}; and by not defining model estimation and inference around a single response
at the expense of the ability to reason about other variables (unlike, e.g., linear regression
models). 

However, information becomes knowledge only when it is placed in context: without it, the
orthodontist has no way to differentiate signal from noise, so the research for better diagnosis
and treatment might be swamped by false positives and false assumptions. Often in everyday
practice the orthodontist's efforts are aimed (or perhaps compelled) to the optimisation
of therapy more than the optimisation of diagnosis. The result is that the therapy is effective,
sometimes extraordinarily effective, but the price of a hasty diagnosis is paid in terms of 
relapse of the pretreatment craniofacial features. The aim of this work is therefore to obtain
an integrated view of the craniofacial features, the treatment and the prognosis to allow
systematic reasoning in the diagnostic process.

The use of Bayesian networks allows us to achieve this aim. We identify the focal morphological
areas of the treatment for Class III malocclusion as the \textit{CoA} segment (the maxillary
length) and the \textit{ANB} angle (the antero-posterior relation of the maxilla to the mandible);
therefore, any apparent effect of the treatment on other cranial features can be disregarded as
noise since it is actually mediated by these two features. Furthermore, by modelling the putative
causal relationships we can study how the effect of an intervention on one feature propagates by
identifying neighbouring features in the DAG and by studying how their distribution changes in
response to various stimuli. We performed such an exercise to validate the consensus Bayesian
network with respect to prior knowledge on malocclusion, with promising results.

To our knowledge, this is the first time a complex system such as craniofacial features has
been modelled in this way with a formal statistical and causal Bayesian network.
The usefulness of such a model is two-fold: it provides an intuitive qualitative description (in
the form of a DAG) of the relationships that link the craniofacial features beyond mere physical
proximity; and it also provides a quantitative description of their behaviour that can be used to
validate the model and to test novel hypotheses by simulation. Collecting clinical data on either
treated or untreated in the context of clinical trials is expensive, time consuming and subject to
many practical, legal and deontological problems. In this context, Bayesian networks provide a way
to perform a preliminary verification of the hypotheses that would be targeted by those trials to
prioritise the trials and allocate resources efficiently\cite{ness}. For instance, in this paper
we identified the focal point of the effect of the facemask therapy. To perform the same task
experimentally without the help of the Bayesian network would require us to check many different
locations; but with the indications provided by the Bayesian networks we can concentrate on 
\textit{CoA} and \textit{ANB} first and possibly avoid further experiments involving the remaining
features. This sequential approach to experimental design and planning is becoming increasingly
common in systems biology to reduce the cost of \textit{in vitro} and \textit{in vivo} research
programmes\cite{pauwels}, and by the pharmaceutical industry to reduce the costs and risk of 
clinical trials\cite{berry}.

While Bayesian networks can deal with the uncertainty in the data, their main limitations lie in
the impact of confounding variables and in the assumptions they make about the distribution of the
features. With a larger number of measurements per patients, for instance, we expect assuming
linear relationships between the features would be a significant limitation, since we would have
enough statistical power to detect nonlinear relationships.

The results of this study show that the Bayesian networks applied to a growing craniofacial 
complex are a useful tool to define a more detailed individualised  prognosis for patients 
affected by the Class III malocclusion, and to mitigate an unpredictable ultimate outcome of
this dysmorphosis.

\section*{Acknowledgments}
%\label{sec:acknowledgments}
GC acknowledges support from EU Projects MULTIPLEX (317532).
%%%%%%%%%%%%%%%%
\section*{Additional information}
\paragraph{Author Contributions}
LF and PA collected the data, GC and MS provided the mathematical framework to analyse such
dataset. All authors contributed equally to the analysis of the dataset and to the
interpretation of the results of this analysis. All authors contributed equally also to the
writing of the manuscript.
\paragraph{Competing financial information}
None of the authors has any financial competing interest.

\newpage
\section{Supplementary Information: Malocclusion Data Analysis}
\thispagestyle{empty}

\subsection{Data}
\begin{figure}
  \centering
  \includegraphics[width=0.4\textwidth]{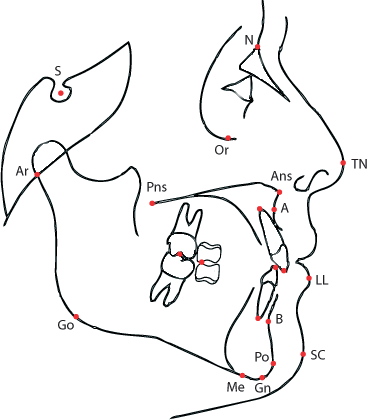}% was {figures/features.png} 
  \caption{Cephalometric landmarks.} 
  \label{features}
\end{figure}

The quantities under consideration are derived from the anatomy of the patient as shown in
Fig.~\ref{features}.

For the data set of $143$ patient, we have the following quantities with two measurements at
ages $T_1$ and $T_2$ (measured in years):
\begin{itemize}
  \item \textit{ID}: anonymised ID code unique to each patient.
  \item \textit{Treatment}: untreated (``NT''), treated with bad results (``TB''), treated with
    good results (``TG'').
  \item \textit{Growth}: a binary variable with values ``Good'' or ``Bad'', determined on the
    basis of CoGn-CoA.
  \item \textit{ANB}: angle between Down's points A and B (degrees). 
  \item \textit{IMPA}: incisor-mandibular plane angle (degrees).
  \item \textit{PPPM}: palatal plane - mandibular plane angle (degrees).
  \item \textit{CoA}: total maxillary length from condilion to Down's point A (mm).
  \item \textit{GoPg}: length of mandibular body from gonion to pogonion (mm).
  \item \textit{CoGo}: length of mandibular ramus from condilion to pogonion (mm).
\end{itemize}

\newpage

%BEGIN FIGURES

\end{document}